# NONSTATIONARY COVARIANCE MODELS FOR GLOBAL DATA[1]

By Mikyoung Jun[2] and Michael L. Stein

*Texas A&M University and University of Chicago*

With the widespread availability of satellite-based instruments, many geophysical processes are measured on a global scale and they often show strong nonstationarity in the covariance structure. In this paper we present a flexible class of parametric covariance models that can capture the nonstationarity in global data, especially strong dependency of covariance structure on latitudes. We apply the Discrete Fourier Transform to data on regular grids, which enables us to calculate the exact likelihood for large data sets. Our covariance model is applied to global total column ozone level data on a given day. We discuss how our covariance model compares with some existing models.

**1. Introduction.** In geophysical and environmental applications, it is common to have spatial data covering a large portion of the Earth. Such data often show nonstationary covariance structure on a global scale. In particular, covariance structures can strongly vary with latitude, although the process may be roughly stationary with respect to longitude.

One example of such data is the Level 3 Total Ozone Mapping Spectrometer (TOMS) data, which gives the daily total column ozone levels. The data are given on a spatially regular grid (1° latitude by 1.25° longitude away from the poles) and there are not many missing observations. A more detailed description of the data is given in Section 4.1. Figure 1(a) shows the standard deviations of the TOMS measurements on May 14, 1990 evaluated at each longitude and (b) shows the similar quantity at each latitude. From

Received February 2008; revised February 2008.

[1]Although the research described in this article has been funded in part by the United States Environmental Protection Agency through STAR cooperative agreement R-82940201-0 to the University of Chicago, it has not been subjected to the Agency's required peer and policy review and therefore does not necessarily reflect the views of the Agency and no official endorsement should be inferred.

[2]Support in part by the National Science Foundation ATM-0620624.

*Key words and phrases.* Nonstationary covariance function, processes on spheres, TOMS ozone data, fast Fourier transform.

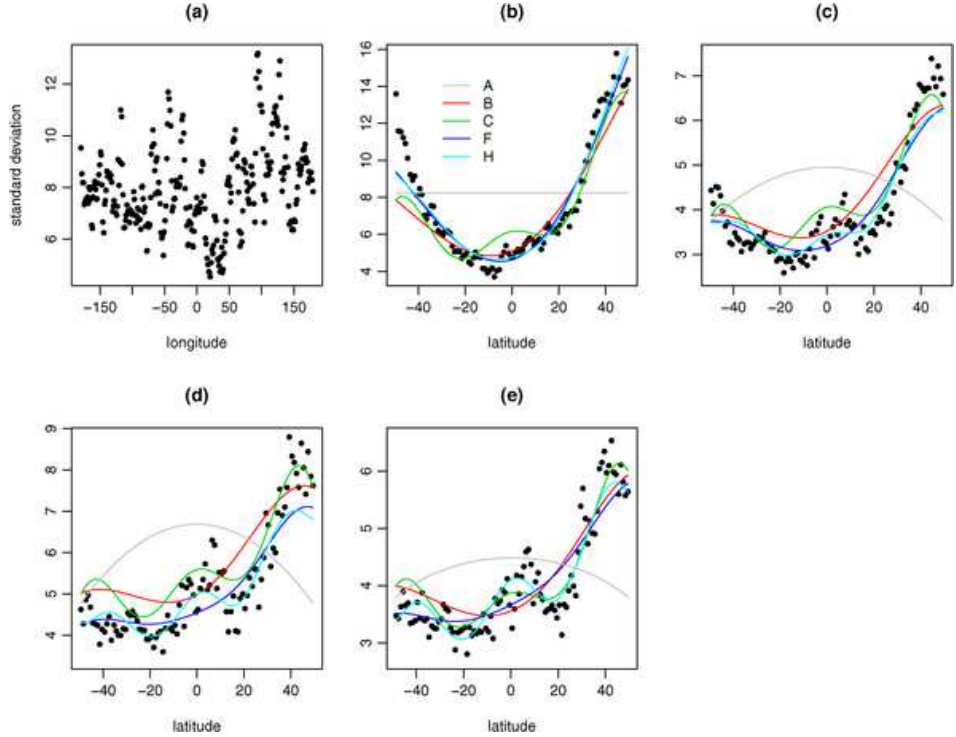

FIG. 1. *Comparison of the empirical (dots) and fitted (colored lines) standard deviations for models A, B, C, F and H. The details on how each quantity are calculated are given in Appendix A. Note that* (a) *does not have fitted values since the quantity plotted is with respect to longitude and is constant under axial symmetry.*

these figures, it is clear that there is a strong dependence of the covariance structure on latitude but not so much dependence on longitude. Table 3 gives the directional variogram values at small scales for a few fixed latitude levels with the same data and the empirical values show strong dependence on latitude.

A number of papers have developed nonstationary spatial covariance functions. Spatial deformations to model nonstationary spatial processes have been used, for example, in Sampson and Guttorp (1992), Perrin and Senoussi (2000), Schmidt and O'Hagan (2003), Clerc and Mallat (2003) and Anderes and Stein (2008). Kernel convolution and its variants have been applied in several papers to create nonstationary covariance functions such as in Higdon (1998), Higdon, Swall and Kern (1999), Fuentes (2002) and Paciorek and Schervish (2006). Nychka, Wikle and Royle (2002) use a wavelet approach to produce nonstationary covariance functions. However, none of these works are suitable for global data since their spatial domain is $\mathbb{R}^d$, not a sphere. Paciorek and Schervish (2006) discuss the possibility of extending their models so that they can be suitable for a spherical domain, but nothing has been explicitly developed or implemented.

There have been several attempts in developing covariance models for global processes. Yaglom (1987) and Gneiting (1999) note that restricting valid isotropic covariance functions in $\mathbb{R}^3$ gives valid covariance functions on a sphere. This idea is simple yet powerful and is able to produce a range of valid covariance functions on a sphere, but only isotropic ones. Another way to generate isotropic covariance functions on spheres is through expansions in spherical harmonics. However, as Jun and Stein (2007) point out, one of the main difficulties in developing covariance functions this way is the mathematical challenge of finding explicit expressions for the covariance function. There are few known closed forms from the spherical harmonics expansion and almost all of these give analytic covariance structures [e.g., Das (2000)], which are too smooth for practical application.

Cressie and Johannesson (2008), Jun and Stein (2007) and Stein (2007, 2008) all consider nonstationary statistical models for total column ozone on a global scale. Cressie and Johannesson (2008) model nonstationary covariance structure on a sphere through a covariance matrix of fixed rank (plus a diagonal matrix), which enables fast computation of the covariance matrices. They allow nonstationarity across latitudes and longitudes. If we only want nonstationarity across latitudes, then we can use what Jones (1963) calls an *axially symmetric* process whose first two moments are invariant to rotations about the Earth's axis. All of the approaches in Jun and Stein (2007) and Stein (2007, 2008) assume that the process is axially symmetric. Similar to Cressie and Johannesson (2008), Stein (2007) uses covariance structures of fixed rank to speed computations, but despite having more than 170 covariance parameters, the resulting model does not do a good job of describing the local behavior of total column ozone. Stein (2008) adds a compactly supported covariance function to this model, which helps to better capture the local fluctuations while maintaining some of the computational advantages of the fixed rank covariance matrices, but still has some problems appropriately describing the nonstationarity across latitudes in the data. The model used in Cressie and Johannesson (2008) has 396 parameters, but it is not clear to us that even this model can accurately capture the local behavior of the process.

The approach described in Jun and Stein (2007) produces a class of rich and flexible nonstationary covariance models for either spatial or spatial–temporal processes with explicit expressions and without requiring nearly as many parameters as approaches based on series expansions. The key idea is to apply differential operators with respect to latitude, longitude and time to a homogeneous process on sphere $\times$ time domain. They give nonrandom functions depending on latitude as coefficients for each differential operator. However, Jun and Stein (2007) do not provide much guidance on how to model the coefficients of the differential operators and their relationship to the resulting covariance structure. The present work seeks to fill this gap by providing some specific approaches for modeling these coefficients of the differential operators.

Whenever we deal with large spatial datasets, we face problems in memory and computation. The covariance functions used here produce covariance matrices that are neither low rank nor sparse, so exact likelihood calculations are not possible with large numbers of irregularly sited observations. One possible solution is to approximate the likelihoods to overcome the computational obstacles [Vecchia (1988), Stein, Chi and Welty (2004), Caragea and Smith (2006), Fuentes (2007)]. For spatially gridded data, however, it is sometimes possible to compute the exact likelihood even for quite large datasets.

In this paper we propose a simple yet flexible way of modeling nonstationary covariance functions on a sphere. We use linear combinations of Legendre polynomials to represent the coefficients of partial differential operators in the approach of Jun and Stein (2007) and find that only a few terms of these polynomials are enough to capture most of the nonstationarity and the local and global variations in the process. Furthermore, by using gridded data and exploiting the structure of the resulting covariance matrix, we achieve great savings in memory and in computation of the likelihood. Our method is applied to global total column ozone levels and the results show that our covariance models with only a modest number of parameters can capture the strong nonstationarity with respect to latitudes as well as the local variation of the process well.

**2. Nonstationary covariance models through differential operators.** For a process $Z(L, l)$ on a sphere ($L$ and $l$ denote latitude and longitude, resp.) such that for a suitable function $K$, $\mathrm{Cov}\{Z(L_1, l_1), Z(L_2, l_2)\} = K(L_1, L_2, l_1 - l_2)$, that is, $Z$ is stationary with respect to longitude, we say the process is axially symmetric [Jones (1963)]. Jun and Stein (2007) propose a class of axially symmetric covariance models for space–time processes that has flexible space–time interactions. For a spatial process, this approach gives a natural way of producing nonstationarity on a purely spatial domain, for instance, different covariance structure along different latitude levels, while retaining the simplification of axial symmetry.

We call a spatial process on a globe homogeneous if its covariance function only depends on the great circle distance (or, equivalently, chordal distance) between two locations. Note that the chordal distance between two locations on a globe, $(L_i, l_i)$, $i = 1, 2$, is given by

$$\mathrm{ch}(L_1, L_2, l_1 - l_2) = 2R\left\{\sin^2\left(\frac{L_1 - L_2}{2}\right) + \cos L_1 \cos L_2 \sin^2\left(\frac{l_1 - l_2}{2}\right)\right\}^{1/2}.$$

Here, $R$ is the radius of the Earth. The great circle distance between the two locations is $\text{gc}(L_1, L_2, l_1 - l_2) = 2R \arcsin\{\text{ch}(L_1, L_2, l_1 - l_2)/(2R)\}$. Let us assume $Z_0$ is homogenous and has mean zero. Then, following Jun and Stein (2007), define a mean zero process $Z$ by

$$Z(L, l) = \left\{ A(L) \frac{\partial}{\partial L} + B(L) \frac{\partial}{\partial l} \right\} Z_0(L, l), \tag{1}$$

where $A$ and $B$ denote nonrandom functions depending on latitude. As explained in Jun and Stein (2007), by making $A$ and $B$ independent of longitude, the resulting process $Z$ is axially symmetric, while it is naturally nonstationary with respect to latitude. Note though that we can also let the functions $A$ and $B$ depend on longitude if we do not want axial symmetry.

Although (1) gives a natural and flexible way of modeling physical processes on a sphere with nonstationary covariance structure, we need to choose the forms for the functions $A$ and $B$ in any specific application. Jun and Stein (2007) modeled ozone levels over a limited latitude band and assumed that $A(L)$ and $B(L)/\cos L$ are constants, which made $\text{Var}\{Z(L, l)\}$ independent of latitude. However, as Figure 1(b) shows, this assumption is inappropriate for total column ozone levels on a global scale.

The covariance models with $A(L)$ and $B(L)/\cos L$ being constants yield constant variance, but it is easy to show that the resulting covariance function is not homogeneous. The covariance models proposed in Jun and Stein (2007), however, do include homogeneous models in some sense by adding a homogeneous process to the field $Z$ in equation (1). That is, by letting

$$Z(L, l) = \left\{ A(L) \frac{\partial}{\partial L} + B(L) \frac{\partial}{\partial l} \right\} Z_0(L, l) + C Z_0(L, l)$$

with $C$ a constant, $Z$ reduces to a homogeneous model when $A(L)$ and $B(L)$ are equal to zero.

Note that correlation values from any homogeneous covariance models in $\mathbb{R}^3$ cannot go below $\inf_{s \geq 0} \frac{\sin s}{s} \approx -0.218$ [Stein (1999)] and Matérn covariance models cannot give negative correlations. However, the differential operators in (1) enable the covariance models to produce negative correlation values down to $-1$. Indeed, it is easy to show that, for any mean square differentiable Matérn model, $\frac{\partial}{\partial l} Z_0(L, l)$ has, as the range parameter tends to infinity, correlation tending to $-1$ for two points at longitudes $180°$ apart.

2.1. *Models for A and B functions.* The approach in (1) may be flexible, but the lack of methodology for selecting and estimating the $A$ and $B$ functions is an impediment to applying the models in real applications to global data. Our idea for modeling $A(L)$ and $B(L)$ is to use linear combinations of Legendre polynomials, which is a class of orthogonal polynomials suited for functions on spheres. Let us consider a linear combination with Legendre polynomials up to order $m$, $P(L; p_0, \ldots, p_m) = \sum_{i=0}^{m} p_i P_i(\sin L)$, where $P_i$ denotes the Legendre polynomial of order $i$. Then, we let $A(L) = P(L; a_0, \ldots, a_p)$ and $B(L) = P(L; b_0, \ldots, b_q)$ for certain integers $p$ and $q$ and constants $a_0, \ldots, a_p, b_0, \ldots, b_q$. We set $a_0 = 1$ to avoid identifiability problems. Specifically, if $a_0$ and $b_0$ are allowed to vary freely, we cannot identify all the parameters, $a_0$, $b_0$ and the sill parameter for $Z_0$. Our goal is, with only a modest number of parameters (if $p = q = 3$, we need 7 parameters for $A$ and $B$) compared to Stein (2007, 2008), to capture nonstationarity at both large and small scales.

We may have to extend the model (1) to make it more flexible. Since the covariance function of $Z$ in (1) involves not only the squared terms of the functions $A$ and $B$ but their cross product terms, once the values of $A$ and $B$ are fixed, $A \cdot B$ is determined completely. Jun and Stein (2007) note this restriction and propose the following extension of (1): for $Z_1, \ldots, Z_N$, i.i.d. copies of $Z_0$, let $Z(L, l) = \sum_{k=1}^{N} \{A_k(L) \frac{\partial}{\partial L} + B_k(L) \frac{\partial}{\partial l}\} Z_k(L, l)$. For example, we may have

$$(2) \quad Z(L, l) = \left\{ A_1(L) \frac{\partial}{\partial L} + B_1(L) \frac{\partial}{\partial l} \right\} Z_1(L, l) + A_2(L) \frac{\partial}{\partial L} Z_2(L, l).$$

This model gives an "interaction" between latitude and longitude through $A_1 \cdot B_1$ that produces a term depending on both magnitude and the sign of longitude lag. Stein (2007) calls a process *longitudinally reversible* if the covariance of the process depends on the longitude lag only through its magnitude. He showed that Level 2 TOMS data is not longitudinally reversible. From the model in (2), the interaction term $A_1 \cdot B_1$ produces covariance models that are not longitudinally reversible. The model in (2) allows further nonstationarity with respect to latitude through $A_2$.

As in Jun and Stein (2007), the process has singularities at the poles unless $\lim_{L \to \pi/2} \{A(L)^2 + B(L)^2 \cos^2 L\} = 0$. The Legendre-polynomial-basis in general does not solve this problem. If we let $A(L) = \sum_{i=0}^{n_1} a_i P_i(\sin L)$ and $B(L) = \sum_{i=0}^{n_2} b_i P_i(\sin L)$, it is straightforward to verify that the condition for the process being mean square continuous at the poles is $\sum_{i=0}^{n_1} a_i = 0$.

**3. Fast computation using the discrete Fourier transform.** In this section we show how to compute the exact full likelihood of data on a grid of 288 longitudes and 100 latitudes. The key idea is the "diagonalization" of the covariance matrix through the Discrete Fourier Transform (DFT) for a block circulant matrix. Note that a block circulant matrix is completely determined by its first block column.

Suppose a process is observed over the full longitude range from $-180°$ to $180°$ and full or partial latitude range. Suppose further that the observed sites are on regular grids, $\{(L_i, l_j) : i = 1, \ldots, m, j = 1, \ldots, n\}$, $|L_i - L_{i-1}| =$ $|L_m - L_1|/(m-1)$, $i = 2, \ldots, m$, and $|l_j - l_{j-1}| = 360/n$ for all $j = 1, \ldots, n$. Let us denote the observed processes

$$\mathbf{Z} = \{Z(L_1, l_1), \ldots, Z(L_m, l_1); \ldots; Z(L_1, l_n), \ldots, Z(L_m, l_n)\}$$

and

$$\widetilde{\mathbf{Z}} = \{Z(L_1, l_1), \ldots, Z(L_1, l_n); \ldots; Z(L_m, l_1), \ldots, Z(L_m, l_n)\}.$$

Notice that $\widetilde{\mathbf{Z}}$ is the same as $\mathbf{Z}$ with the elements rearranged. Now, if the process $Z$ is axially symmetric, it is easy to see that the covariance matrix of $\mathbf{Z}$ is block circulant. Or, the covariance matrix of $\tilde{\mathbf{Z}}$ is an $n \times n$ block matrix with $m \times m$ circulant blocks. Theorems 5.6.4 and 5.7.2 in Davis (1979) state that, through the DFT, the covariance matrix of $\mathbf{Z}$ can be diagonalized and become a block diagonal matrix; the same diagonalization holds for the covariance matrix of $\widetilde{\mathbf{Z}}$ after rearranging the rows and columns. In particular, if we let $\mathbf{Z}_L = \{Z(L, l_1), \ldots, Z(L, l_n)\}$ for any latitude $L$ and write $F$ for the DFT operator, the covariance matrix of $\{F\mathbf{Z}_{L_1}, \ldots, F\mathbf{Z}_{L_m}\}$ consists of $n \times n$ diagonal matrices as its blocks. Then it is easy to see that the covariance matrix of $F\mathbf{Z}$ should be a block diagonal matrix whose block size is $m \times m$, since it is just the covariance matrix of $\{F\mathbf{Z}_{L_1}, \ldots, F\mathbf{Z}_{L_m}\}$ with rows and columns rearranged.

Now, instead of directly calculating the likelihood of $\mathbf{Z}$, we calculate the likelihood of $F\mathbf{Z}$ whose covariance matrix is a block diagonal matrix. Hence, instead of computing the Cholesky decomposition of a $mn \times mn$ matrix, we only need to decompose $n$ matrices of size $m \times m$ matrix. As long as the data are on regular grids with full longitude range and no missing data, the computation of the exact likelihood is very fast using the fast Fourier transform. Note that in calculating loglikelihood, we need $m$ DFTs and $n$ Cholesky decompositions of matrices of size $m \times m$. Thus, the total computational complexity for DFT is $O(mn \log n)$ and that for the Cholesky decomposition is $O(m^3 n)$. In terms of memory, since circulant matrices are determined by their first column, we do not have to save the entire covariance matrix. This reduces the size of required memory substantially.

**4. Application: TOMS Level 3 data.** We now show the application of the covariance models to total column ozone level data on a global scale.

4.1. *Data.* During the period of November 1, 1978 to May 6, 1993, Nimbus 7 carried a TOMS instrument and the data from this satellite are either in Level 2 or Level 3 versions. TOMS Level 2 data have been analyzed in some recent papers in the statistical literature, including Cressie and Johannesson (2008) and Stein (2007, 2008). Level 2 data give spatially and temporally irregular measurements following the satellite scanning tracks (measurements are 8 seconds apart) and there is a significant amount of missing observations. TOMS Level 3 data are post processed from Level 2 data and they are on regular grids [1 degree latitude by 1.25 degrees longitude for pixels with latitude from 50° S to 50° N, see Krueger et al. (1998) for more details] as daily averages. TOMS Level 3 data are obtained from an ad hoc method to average Level 2 data pixel by pixel. Both Levels 2 and 3 data, along with more information on the data, can be obtained from http://toms.gsfc.nasa.gov/ozone/ozone_v8.html. Cressie and Johannesson (2008) produce new Level 3 data through statistical models rather than ad hoc averaging. Although there is loss of information in Level 3 data, especially fine scale spatial and temporal variations, data on grids with global coverage and few missing observations are convenient to focus on the study of the covariance structure of the process purely due to the computational efficiencies described in Section 3. Our covariance models are not restricted to data on regular grids, but exact likelihood computations would then require much more memory and computation.

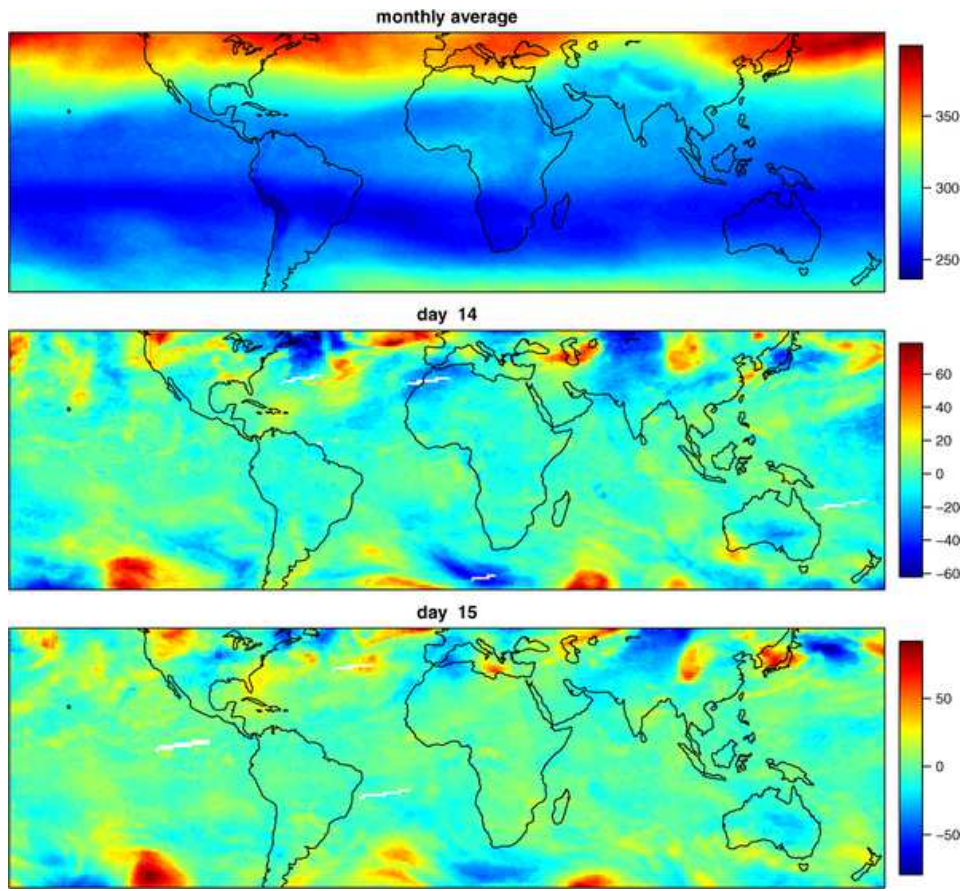

FIG. 2. *Monthly average and daily data minus monthly average of TOMS ozone Level 3 data. Top figure is for monthly average of May 1990, middle for May 14 1990 and the bottom for May 15 1990.*

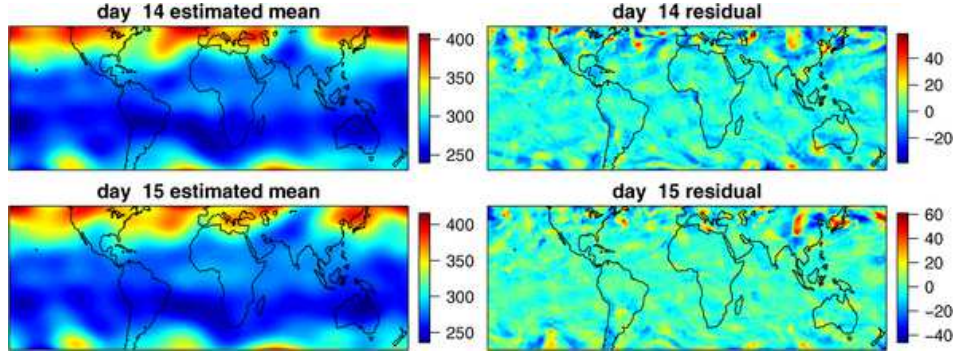

FIG. 3. *Estimated mean structure of the data for May 14 and May 15, 1990, using spherical harmonics ($\{Y_n^m(\sin L, l)|n = 0, 1, 2, \ldots, m = -n, \ldots, n\}$ for $n = 12$, see Section 4.2), and the residuals (data minus estimated mean).*

We use the original NASA-produced TOMS Level 3 data (Level 2 data aggregated through pixel by pixel averaging) for May 14–15, 1990. There are 288 longitude points (evenly spaced over $-180°$ to $180°$) and 100 latitude points (evenly spaced over $50°$ S–$50°$ N). The data from latitude levels beyond this range have fewer than 288 observations per latitude. The size of the data is thus 28,800 per day. There are still a few pixels with missing data (e.g., out of 28,800 pixels 62 pixels for May 14 and 71 for May 15) and we naively impute these locations. That is, the missing pixels are filled with the average of available data from the 8 neighboring cells. Considering the small fraction of missing observations, the method of imputation should not affect the fitted results significantly. In fact, we tried various other imputation methods and the results do not change much.

4.2. *Modeling the mean structure.* Since we focus on modeling the covariance structure of this data, we should somehow filter the data and make the process close to mean zero. We first tried to subtract the monthly average of May from the data for each day in May. After this filtering, we would expect the residual should be closer to mean zero and Gaussian. However, as Figure 2 shows, the residuals have noticeable patterns. For example, for both May 14 and 15, there are hot spots in the high and low latitude area. While these phenomena could be explained by Rossby waves [Holton (1992)], our covariance models developed here do not have the ability to capture such patterns and so we seek to model these patterns with the mean function.

Spherical harmonics provide a natural basis for capturing the large-scale patterns displayed in Figure 2. This is similar to what has been done in Stein (2007, 2008). Specifically, we regress the ozone levels with $\{Y_n^m(\sin L, l)|n = 0, 1, 2, \ldots, m = -n, \ldots, n\}$ for $n = 12$. Figure 3 shows the estimated mean on May 14–15, 1990 and the residuals. Note that the spherical harmonics terms capture most of the patterns in the mean and the residuals do not have the noticeable patterns that we see in Figure 2. We tried a few different values of $n$ around 12 and the filtered results are not sensitive to the choice of $n$.

4.3. *Tapering of the data.* An issue with the TOMS Level 3 data is the discontinuity at the international date line (IDL) [Fang and Stein (1998)]. TOMS Level 3 data are post processed data from the original satellite scans and they give daily averages for the full range of longitudes. Therefore, there are substantial discrepancies in the data at the IDL, in particular, at longitudes $\pm 179.375°$. However, since our model is purely spatial and axially symmetric, it cannot model a noticeable discrepancy at the IDL.

A natural way of dealing with this issue is data tapering. To explore the effect of tapering, at each latitude level, we tapered the data by a split cosine-bell taper [Tukey (1967)] with 5% of each end of the data tapered. Note though that our data are actually on a circle. Therefore, the taper set the ozone levels at longitudes $\pm 180°$ to be zero and the data at longitude ranges (approximately) $-180°$ to $-163°$ and $163°$ to $180°$ are tapered accordingly. We do not taper the original data but the residual after subtracting the mean estimated in Section 4.2.

Note that in this case the process is not axially symmetric anymore and, thus, the covariance matrix is no longer block circulant, but only block Toeplitz. Thus, the covariance matrix of $F\mathbf{Z}$ is block diagonal only asymptotically. We tried fitting the model to the tapered data in two different scenarios. First, we simply ignored the fact that the process is no longer axially symmetric and performed the computation as in Section 3. A second approach is that we approximated the covariance matrix of $F\mathbf{Z}^t$ ($\mathbf{Z}^t = \{Z^t(L_1, l_1), \ldots, Z^t(L_m, l_1); \ldots; Z^t(L_1, l_n), \ldots, Z^t(L_m, l_n)\}$) by a block diagonal matrix, but we modified each block in the diagonal so that the block diagonals give the covariances that take account of the tapering. We applied both approaches for more than 20 days of May, 1990 (each day fit separately) and found that the parameter estimates for the two approaches are quite similar. Therefore, for computational simplicity, we decided to apply the first approach ignoring the effect of tapering in fitting the covariance functions.

4.4. *Covariance models.* We compare the nonstationary models developed in Section 2.1 along with a few other nonstationary covariance models. The covariance models that we compare here are axially symmetric and in nested format.

TABLE 1
*Covariance models and the increase of maximized loglikelihood values (denoted by $\Delta$ loglikelihood) compared to the maximized loglikelihood value for A, which is $-128568.9$*

| Model | A | B | C | D | E | F | G | H | I | J |
|---|---|---|---|---|---|---|---|---|---|---|
| $m$ | 0 | 3 | 6 | 0 | 0 | 3 | 3 | 3 | 6 | 6 |
| $n_1$ | — | — | — | 3 | 6 | 3 | 6 | 6 | 6 | 6 |
| $n_2$ | — | — | — | 3 | 6 | 3 | 0 | 6 | 6 | 6 |
| $n_3$ | — | — | — | — | — | — | — | — | — | 6 |
| # parameters | 4 | 7 | 10 | 14 | 20 | 17 | 16 | 23 | 26 | 30 |
| $\Delta$ loglikelihood | 0 | 1764.1 | 1900.1 | 1867.6 | 2052.3 | 2671.8 | 2714.6 | 2803.4 | 2871.6 | 2914.3 |

Suppose $Z_0$ in (1) has mean zero and a homogeneous covariance function $K_0$. In particular, let $K_0$ be a Matérn covariance function when viewed as a covariance function on $\mathbb{R}^3$:

$$(3) \quad \text{Cov}\{Z_0(L_1, l_1), Z_0(L_2, l_2)\} = K_0(d; \alpha, \beta, \nu) = \alpha(d/\beta)^\nu \mathcal{K}_\nu(d/\beta),$$

where $\mathcal{K}_\nu$ is the modified Bessel function of the third kind of order $\nu$ [Stein (1999)] and $d = \text{ch}(L_1, L_2, l_1 - l_2)$. Note that $\beta$ is the spatial range parameter. Our simplest model for $Z$ is a rescaled version of $Z_0$ with a nugget effect: $Z(L, l) = P(L; k_0, \ldots, k_m) Z_0(L, l)$ plus a nugget effect, yielding the covariance function

$$(4) \quad \begin{aligned} &K_1(L_1, L_2, l; \alpha, \beta, \nu, \varepsilon, k_0, \ldots, k_m) \\ &= P(L_1; k_0, \ldots, k_m) P(L_2; k_0, \ldots, k_m) K_0(d; \alpha, \beta, \nu) \\ &\quad + \varepsilon \cdot \mathbf{1}_{(L_1 - L_2 = l = 0)}, \end{aligned}$$

for $d = \text{ch}(L_1, L_2, l)$. By multiplying $Z_0$ by the function $P(L; k_0, \ldots, k_m)$, $K_1$ can capture the strong dependence of variance on latitude through a fairly simple model. To avoid identifiability problems, we set $k_0 = 1$. When $m = 0$, the variance of the process is constant with respect to latitude and the process is isotropic.

Next, we add the component developed in Section 2.1 to $K_1$. That is,

$$(5) \quad \begin{aligned} &K_2(L_1, L_2, l; \alpha, \beta, \nu, \varepsilon, k_0, \ldots, k_m, \alpha_1, \beta_1, \nu_1, a_0, \ldots, a_{n_1}, b_0, \ldots, b_{n_2}) \\ &= K_Z(L_1, L_2, l; \alpha_1, \beta_1, \nu_1, a_0, \ldots, a_{n_1}, b_0, \ldots, b_{n_2}) \\ &\quad + K_1(L_1, L_2, l; \alpha, \beta, \nu, \varepsilon, k_0, \ldots, k_m). \end{aligned}$$

Here, $K_Z$ in (5) is the covariance function of (1) with $\alpha_1$, $\beta_1$ and $\nu_1$ being the parameters for (3), $A(L) = P(L; a_0, \ldots, a_{n_1})$ and $B(L) = P(L; b_0, \ldots, b_{n_2})$. To avoid identifiability problems, we set $k_0 = a_0 = 1$. We also have an extended version of $K_2$, denoted by $K'_Z$, obtained by using $Z$ as in (2) for the process $Z$ in (5). That is, $K'_Z(L_1, L_2, l; \alpha_1, \beta_1, \nu_1, a_0, \ldots, a_{n_1}, b_0, \ldots, b_{n_2}, c_0, \ldots,$ $c_{n_3}) = K_Z(L_1, L_2, l; \alpha_1, \beta_1, \nu_1, a_0, \ldots, a_{n_1}, b_0, \ldots, b_{n_2}) + K_Z(L_1, L_2, l; \alpha_1, \beta_1, \nu_1, c_0, \ldots, c_{n_3})$ with $A_2(L) = P(L; c_0, \ldots, c_{n_3})$. A summary of the covariance models with their numbers of parameters is given in Table 1.

4.5. *Fitted results.* The full likelihoods are calculated using the discrete Fourier transform as described in Section 3; we use the R functions *nlm* and *optim* to maximize the likelihoods. We tried several initial points for each optimization and found that all of the optimizations yielded the same maximum points. Table 1 gives the maximized loglikelihood values for each covariance model. First notice that the rescaled isotropic model with 3 additional parameters (model B) compared to the simplest isotropic model (model A) increases the loglikelihood by over 1700. This large increase is expected since there is clear latitude dependency of variance as shown in Figure 1(b). By adding the $K_Z$ function in (5) to the rescaled isotropic model, we get an additional substantial increase of the loglikelihood. That is, the loglikelihood of model F increases by about 907 and that of model G increases about 950 from the loglikelihood of model B. Models D and E, which have a $K_Z$ component but not the rescaled version of the isotropic part in $K_1$, do not fit the data as well as models F or G, despite model E having more parameters than models F and G. Model J, which has covariance function $K'_Z$ as described below (5), does not improve the fit dramatically over model G.

The estimated covariance parameters (MLE) for some of the models in Table 1 along with their asymptotic standard errors are presented in Table 2. The unit of spatial distance is km. The asymptotic standard errors are given by the square roots of the diagonal elements of the inverse Hessian matrix, evaluated at the MLE values. For model H, even though the optimization by the R functions *nlm* and *optim* claimed that it reached the maximum point, the Hessian matrix was almost singular. This is due to two parameters, $\alpha_1$ and $\beta_1$, for which the data provide relatively little information compared to the other parameters. Figure 4 shows the shape of profile loglikelihood for model H. The profile loglikelihood is indeed flat for parameters $\beta_1$ and $\nu_1$ compared to parameters $\beta$ and $\nu$, respectively. The two parameters, $\alpha$ and $\alpha_1$, are not comparable in magnitude and, thus, direct comparison of the shape of the loglikelihood from this figure is not appropriate. Because of this problem with model H, we fixed the values for the two parameters, $\alpha_1$ and $\beta_1$, from the first optimization and ran the optimization again for the other parameters. The asymptotic standard errors given for model H are from the second optimization. Notice first that the MLE values of the sill, spatial range and smoothness parameters for the three models do not vary much. The estimated nugget is somewhat bigger for model B compared to the other two models. One of the big differences between model B and models F and H is that models F and H have the differential operators terms while model B does not. Model B simply has spatially varying variance terms but its correlation structure is homogeneous. Therefore, some of the nonstationarity in the data is not accounted for in model B and we get larger estimated values for the nugget effect. Some of the $a_i$'s and $b_i$'s have rather large asymptotic standard errors, which may be a sign that the data do not provide enough information to estimate these particular parameters well. From Figure 4, it is interesting to note that the profile loglikelihood has two local modes with respect to $a_2$ for model H.

TABLE 2
*Maximum likelihood estimates of covariance parameters for May 14, 1990 along with their asymptotic standard errors. For model H, standard errors are based on treating $\alpha_1$ and $\beta_1$ as known (see Section 4.5 for details)*

|       | B             | F                     | H               |
|-------|---------------|-----------------------|-----------------|
| $\alpha$ | 64.89 (2.64)  | 73.59 (3.81)          | 73.98 (3.99)    |
| $\beta$  | 218.65 (8.88) | 260.20 (11.54)        | 262.88 (11.82)  |
| $\nu$    | 1.20 (0.038)  | 1.23 (0.040)          | 1.26 (0.040)    |
| $\varepsilon$ | 1.76 (0.089) | 0.41 (0.11)        | 0.49 (0.11)     |
| $k_1$ | 0.48 (0.023)  | 0.46 (0.034)          | 0.51 (0.035)    |
| $k_2$ | 0.81 (0.023)  | 1.061 (0.024)         | 1.08 (0.024)    |
| $k_3$ | 0.071 (0.030) | 0.15 (0.041)          | 0.22 (0.044)    |
| $\alpha_1$ | —        | 6.13e–05 (6.03e–06)   | 4.81e–06 (—)    |
| $\beta_1$  | —        | 53.14 (1.32)          | 58.65 (—)       |
| $\nu_1$    | —        | 2.5 (2.11e–04)        | 2.25 (1.51)     |
| $a_1$ | —             | 0.34 (0.069)          | 2.36 (0.84)     |
| $a_2$ | —             | −0.034 (0.061)        | −15.65 (0.89)   |
| $a_3$ | —             | −0.045 (0.078)        | 0.62 (1.22)     |
| $a_4$ | —             | —                     | −17.13 (0.88)   |
| $a_5$ | —             | —                     | −0.11 (0.69)    |
| $a_6$ | —             | —                     | −10.42 (0.53)   |
| $b_0$ | —             | 0.15 (0.036)          | 3.07 (1.44)     |
| $b_1$ | —             | 0.57 (0.37)           | −7.18 (2.64)    |
| $b_2$ | —             | 0.89 (0.034)          | 13.91 (0.52)    |
| $b_3$ | —             | 0.27 (0.033)          | −13.14 (4.33)   |
| $b_4$ | —             | —                     | 10.90 (0.39)    |
| $b_5$ | —             | —                     | −7.55 (2.39)    |
| $b_6$ | —             | —                     | 6.66 (0.62)     |

Figure 1(b) shows the comparison of the empirical and fitted values for the standard deviations, $K(L, L, 0)^{1/2}$, with respect to $L$. The covariance model A naturally gives constant standard deviations. The other models give similar fits except for model C, which creates a spurious pattern around the equator. This spurious pattern may be a sign that we need bigger $m$ for model C since the model is without the asymmetric term and is thus less flexible. In Figure 1(c)–(e), it is clear that the fit to the data improves significantly by adding the asymmetric term as in $K_2$. Model H captures the patterns near the equator in the data quite well in spite of the problem in convergence of the optimization mentioned in the previous paragraph. Model C gives similar patterns as model H, but it mostly overestimates variation throughout all latitude levels.

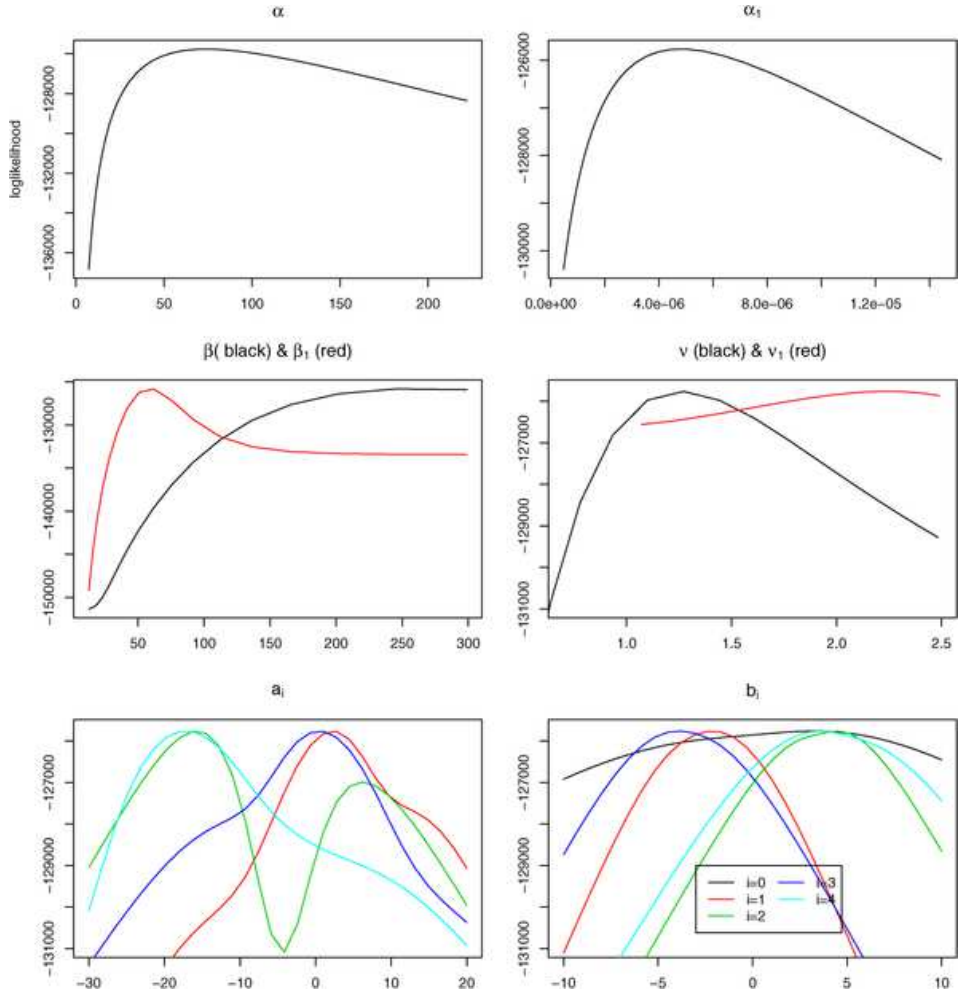

Fig. 4. *Shape of profile loglikelihood for model H. Top: profile loglikelihood with respect to $\alpha$ (left) and $\alpha_1$ (right). Middle: profile loglikelihood with respect to $\beta$ and $\nu$ (black) and $\beta_1$ and $\nu_1$ (red). Bottom: profile loglikelihood with respect to $a_i$'s and $b_i$'s for $i=0$ (only for $b_0$) and $i=1,\ldots,4$.*

Table 3 gives the directional variogram values for the two nearest points in various directions and the corresponding fitted values. In comparing values across latitudes, we have to be careful since the distance that corresponds to a degree longitude changes with latitudes. The distance between neighboring two cells in N–S direction is always the same distance apart across latitudes. Most of the models give good fits to the empirical values and model H overall gives the best fit. Notice that model A gives a symmetric fit for both Northern and Southern Hemisphere due to its limitation. The empirical values in the table (compare NW and NE at 40° N or SE and SW at 40° S) suggest that the process is not longitudinally reversible, that is, $K(L_1, L_2, l) \neq K(L_1, L_2, -l)$ for some $L_1, L_2$ and $l$, which coincides with the finding of Stein (2007) for TOMS Level 2 data. Our fitted models do not capture this moderate local anisotropy, despite the fact that some of the models allow for longitudinal irreversibility.

TABLE 3
*The square root of the directional variograms at fixed latitudes. The details on how the quantities are calculated are given in Appendix B*

|   | Latitude | SE | S | SW | W | Latitude | NW | N | NE | E |
|---|---|---|---|---|---|---|---|---|---|---|
| Empirical |         | 4.03 | 3.53 | 3.99 | 3.23 |         | 3.66 | 3.18 | 3.79 | 3.12 |
| A         |         | 5.74 | 4.30 | 5.74 | 4.95 |         | 5.74 | 4.30 | 5.74 | 4.95 |
| B         | 0.5° S  | 3.97 | 3.13 | 3.97 | 3.51 | 0.5° N  | 4.04 | 3.18 | 4.04 | 3.54 |
| C         |         | 4.65 | 3.57 | 4.65 | 4.05 |         | 4.68 | 3.59 | 4.68 | 4.07 |
| F         |         | 3.67 | 3.34 | 3.96 | 3.18 |         | 4.01 | 3.39 | 3.72 | 3.20 |
| H         |         | 3.94 | 3.71 | 4.33 | 3.44 |         | 4.39 | 3.75 | 3.98 | 3.46 |
| Empirical |         | 3.55 | 3.23 | 3.81 | 2.95 |         | 4.29 | 3.38 | 3.73 | 2.98 |
| A         |         | 5.61 | 4.30 | 5.61 | 4.75 |         | 5.61 | 4.30 | 5.61 | 4.75 |
| B         | 20.5° S | 3.99 | 3.20 | 3.99 | 3.45 | 20.5° N | 5.51 | 4.25 | 5.51 | 4.63 |
| C         |         | 3.60 | 2.91 | 3.60 | 3.15 |         | 4.73 | 3.70 | 4.73 | 4.01 |
| F         |         | 3.84 | 3.31 | 4.01 | 3.18 |         | 5.20 | 4.35 | 5.17 | 4.12 |
| H         |         | 3.67 | 3.05 | 3.77 | 3.02 |         | 4.77 | 3.93 | 4.85 | 3.84 |
| Empirical |         | 5.08 | 5.15 | 6.04 | 3.21 |         | 9.21 | 7.07 | 7.86 | 6.48 |
| A         |         | 5.25 | 4.30 | 5.25 | 4.16 |         | 5.25 | 4.30 | 5.25 | 4.16 |
| B         | 40.5° S | 4.79 | 3.98 | 4.79 | 3.84 | 40.5° N | 7.84 | 6.33 | 7.84 | 6.06 |
| C         |         | 5.10 | 4.21 | 5.10 | 4.04 |         | 8.28 | 6.70 | 8.28 | 6.39 |
| F         |         | 4.93 | 4.11 | 4.94 | 3.67 |         | 7.72 | 6.51 | 8.00 | 5.82 |
| H         |         | 5.05 | 4.31 | 5.10 | 3.75 |         | 8.06 | 6.88 | 8.22 | 5.92 |
| Empirical |         | 5.99 | 4.44 | 5.57 | 4.52 |         | 9.03 | 7.25 | 8.89 | 6.48 |
| A         |         | 5.10 | 4.30 | 5.10 | 3.90 |         | 5.10 | 4.30 | 5.10 | 3.90 |
| B         | 46.5° S | 5.05 | 4.28 | 5.05 | 3.88 | 46.5° N | 8.50 | 7.05 | 8.50 | 6.29 |
| C         |         | 5.29 | 4.47 | 5.29 | 4.07 |         | 8.74 | 7.27 | 8.74 | 6.53 |
| F         |         | 5.26 | 4.45 | 5.25 | 3.74 |         | 8.53 | 7.32 | 8.85 | 6.17 |
| H         |         | 5.23 | 4.35 | 5.08 | 3.71 |         | 8.87 | 7.50 | 8.77 | 6.16 |

The shapes of the fitted functions $A$ and $B$ as well as $P(L; k_0, \ldots, k_m)$ [in (4)] for each model are given in Figure 5. The shape of $P(L; k_0, \ldots, k_m)$ resembles the shape of the empirical variances in Figure 1(b). For most latitude values, the magnitude of $B$ is close to zero compared to that of

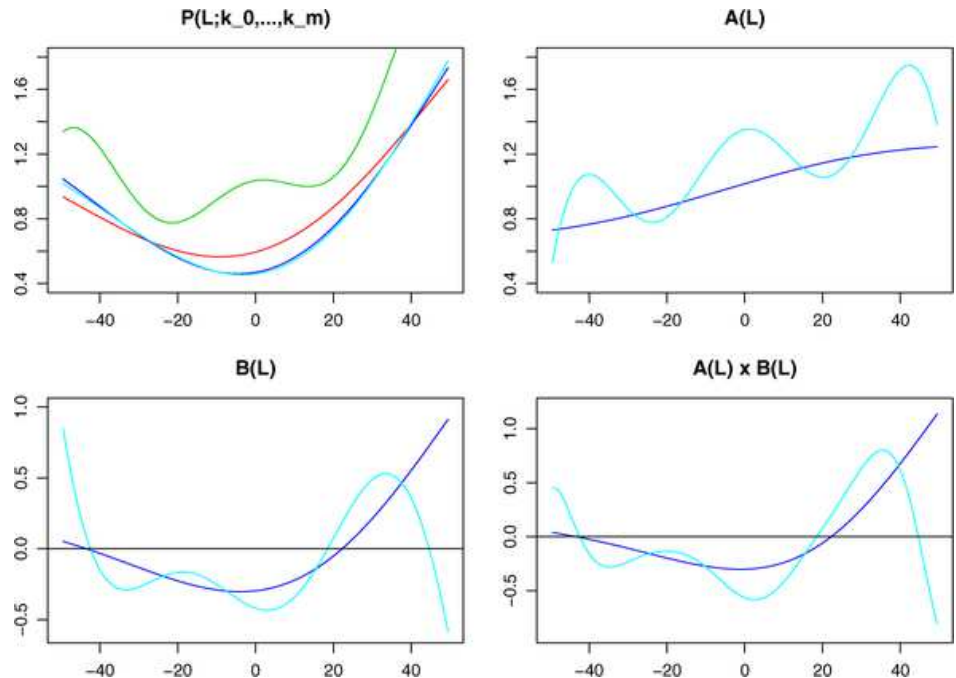

Fig. 5. *The estimated values of $P(L; k_0, \ldots, k_m)$, $A(L)$, $B(L)$ and $A(L)B(L)$ with respect to latitude for the covariance models in Table 1. The color legend is the same as in Figure 1.*

A. In fact, the fit with $B$ set to be zero (model G) is not much worse than model H (see Table 1 for loglikelihood values). For model G, most of the other parameter values stay nearly the same as in model H except the smoothness parameter $\nu_1$, which is close to the boundary of the parameter space (note $\nu_1 > 1$). This may be a sign of model misfit.

**5. Discussion.** In the paper we rather informally compare different models using several criterion such as maximized likelihood values or the comparison between empirical data and fitted values. We would need more formal criteria such as AIC or BIC that penalizes the number of parameters if we want to perform formal model selection.

We arbitrarily choose the values of $m, n_1$ and $n_2$ and do not estimate them. The point is that with moderate values of these constants, we achieve great flexibility of the covariance structure. As in Tables 1 and 2 and Figures 1 and 5, the performance/behavior of the estimated covariances do not change much for models F through J. The models with $m = 2$ or 4 should not give much different covariance structure than $m = 3$. In choosing these constants, there is a trade off between the flexibility of the covariance structure and the simplicity of the covariance models (in terms of number of parameters).

We demonstrate that the characteristics of nonstationarity in our data are similar to those in Stein (2007) that deal with Level 2 data (ungridded version). However, it is possible that some of the nonstationarity come from the gridding of Level 3 data and this effect would likely be worse if we went beyond our latitude range here.

A Bayesian approach may be a natural alternative for parameter estimation and imputation of missing data. The computational technique in Section 3 requires complete observations and we used a simple imputation technique to fill in missing values. For applications with a nonnegligible amount of missing data, a Bayesian framework, if computationally feasible, may be a better way of imputing the missing values as well as estimating the covariance parameters and their uncertainty.

The nonstationary covariance models developed in this paper have a very natural extension to spatial-temporal processes. By introducing a differential operator with respect to time in addition to those with respect to longitude and latitude, the models should not only capture spatial nonstationary patterns, but also create flexible space–time interaction, such as space–time asymmetry. Jun and Stein (2007) demonstrate the flexibility of this approach in capturing space–time asymmetry in total column ozone measurements in the Northern hemisphere, but they only considered a limited latitude range and did not allow sufficient flexibility in their models to fit the nonstationarity in the spatial variations of total column ozone on a global scale. The idea of applying linear combinations of Legendre polynomials for the coefficients of the differential operators introduced in this paper can also be applied to obtain flexible space–time interactions, for example, allowing the speed of flow of ozone across longitudes to vary with latitude.

## APPENDIX A

This appendix describes how each quantity in Figure 1 is calculated. We have 100 equally spaced latitude points from $-49.5°$ to $49.5°$ and we denote these $L_1, \ldots, L_{100}$. We also have 288 equally spaced longitude points from $-179.375°$ to $179.375°$ and we denote these $l_1, \ldots, l_{288}$. Let us also denote $\bar{Z}^L(\cdot) = \frac{1}{100} \sum_{i=1}^{100} Z(L_i, \cdot)$ and $\bar{Z}^l(\cdot) = \frac{1}{288} \sum_{i=1}^{288} Z(\cdot, l_i)$. We let $\Delta L = 1°$ and $\Delta l = 1.25°$. Then:

(a) $\operatorname{Var}\{Z(\cdot, l)\} = \frac{1}{99} \sum_{i=1}^{100} \{Z(L_i, l) - \bar{Z}^L(l)\}^2$ for each $l$.
(b) $\operatorname{Var}\{Z(L, \cdot)\} = \frac{1}{287} \sum_{i=1}^{288} \{Z(L, l_i) - \bar{Z}^l(L)\}^2$ for each $L$.
(c) For each $L$,

$$\operatorname{Var}\{Z(L, l) - Z(L, l - \Delta l)\}$$
$$= \frac{1}{286} \sum_{i=2}^{288} \left[ Z(L, l_i) - Z(L, l_{i-1}) - \frac{1}{287} \sum_{j=2}^{288} \{Z(L, l_j) - Z(L, l_{j-1})\} \right]^2.$$

(d) For each $L$,

$$\text{Var}\{Z(L, l+\Delta l) - 2Z(L,l) + Z(L, l-\Delta l)\}$$
$$= \frac{1}{285} \sum_{i=2}^{287} \left[ Z(L, l_{i+1}) - 2Z(L,l) + Z(L, l_{i-1}) \right.$$
$$\left. - \frac{1}{286} \sum_{j=2}^{287} \{Z(L, l_{j+1}) - 2Z(L,l) + Z(L, l_{j-1})\} \right]^2.$$

(e) For each $L_i$, $i = 2, \ldots, 100$,

$$\text{Var}\{Z(L_i, l+\Delta l) - Z(L_i, l) - Z(L_{i-1}, l+\Delta l) + Z(L_{i-1}, l)\}$$
$$= \frac{1}{286} \sum_{j=1}^{287} \left[ Z(L_i, l_{j+1}) - Z(L_i, l_j) - Z(L_{i-1}, l_{j+1}) + Z(L_{i-1}, l_j) \right.$$
$$- \frac{1}{287} \sum_{k=1}^{287} \{Z(L_i, l_{k+1})$$
$$\left. - Z(L_i, l_k) - Z(L_{i-1}, l_{k+1}) + Z(L_{i-1}, l_k)\} \right]^2.$$

## APPENDIX B

This appendix gives how quantities in Table 3 are calculated. Suppose the fixed latitude level is $L_i$ for some $i = 2, \ldots, 99$. Then the empirical variogram for SE direction is given by $\frac{1}{287} \sum_{j=1}^{287} \{Z(L_i, l_j) - Z(L_{i-1}, l_{j+1})\}^2$ and the corresponding fitted value is $K(L_i, L_i, 0) + K(L_{i-1}, L_{i-1}, 0) - 2K(L_i, L_{i-1}, l_j - l_{j+1})$ for $K$, an estimated covariance function of $Z$. Similarly, the empirical variogram for S direction is given by $\frac{1}{288} \sum_{j=1}^{288} \{Z(L_i, l_j) - Z(L_{i-1}, l_j)\}^2$ and the corresponding fitted value is $K(L_i, L_i, 0) + K(L_{i-1}, L_{i-1}, 0) - 2K(L_i, L_{i-1}, 0)$. The values for the other directions are calculated similarly.

**Acknowledgments.** The authors are grateful to the Associate Editor and two anonymous referees for helpful comments and suggestions.

DEPARTMENT OF STATISTICS  
TEXAS A&M UNIVERSITY  
3143 TAMU  
COLLEGE STATION, TEXAS 77843-3143  
USA  
E-MAIL: mjun@stat.tamu.edu

DEPARTMENT OF STATISTICS  
UNIVERSITY OF CHICAGO  
5734 S. UNIVERSITY AVENUE  
CHICAGO, ILLINOIS 60637  
USA  
E-MAIL: stein@galton.uchicago.edu